\documentclass[12pt]{article}
\usepackage{graphicx}

\newcommand\pubnumber{DPF2015-91}
\newcommand\pubdate{\today}

\def\Title#1{\begin{center} {\Large #1 } \end{center}}
\def\Author#1{\begin{center}{ \sc #1} \end{center}}
\def\Address#1{\begin{center}{ \it #1} \end{center}}

\newcommand\pubblock{\rightline{\begin{tabular}{l} \pubnumber\\
         \pubdate  \end{tabular}}}
\newenvironment{Abstract}{\begin{quotation}  }{\end{quotation}}
\newenvironment{Presented}{\begin{quotation} \begin{center} 
             PRESENTED AT\end{center}\bigskip 
      \begin{center}\begin{large}}{\end{large}\end{center} \end{quotation}}

\def\beq{\begin{equation}}
\def\eeq#1{\label{#1}\end{equation}}
\def\eeqn{\end{equation}}

\def\beqa{\begin{eqnarray}}
\def\eeqa#1{\label{#1}\end{eqnarray}}
\def\eeqan{\end{eqnarray}}

\let\bar=\overbar

\def\Dslash{\not{\hbox{\kern-4pt $D$}}}
\def\dslash{\not{\hbox{\kern-2pt $\del$}}}

\def\msb{{\bar{\ssstyle M \kern -1pt S}}}

\usepackage{amsmath}
\usepackage{cite}
\usepackage{subfig}
\usepackage{graphicx}
\begin{document}
\begin{titlepage}
\pubblock

\vfill
\Title{Multiboson measurements and limits on Anomalous Gauge Couplings from the CMS Experiment}
\vfill
\Author{Joshua Milo Kunkle}
\Address{University of Maryland}
\vfill
\begin{Abstract}
Recent measurements of multiboson production from the CMS experiment will be presented, as well as limits on anomalous triple and quartic gauge couplings.  Precision measurements of multiboson production allow a basic test of the Standard Model, where higher order QCD and electroweak corrections can be probed.  In addition searches of physics beyond the Standard Model in multiboson final states rely on precise determination of the Standard Model multiboson processes.  The presence of triple and quartic gauge couplings in multiboson production also allows for tests of modification of these vertices from new physics.  Prospects for future measurements will also be shown.  With the increased center of mass energy of the LHC and the integrated luminosity that will be collected in LHC Run 2, the limits on anomalous gauge couplings will improve significantly.
\end{Abstract}
\vfill
\begin{Presented}
DPF 2015\\
The Meeting of the American Physical Society\\
Division of Particles and Fields\\
Ann Arbor, Michigan, August 4--8, 2015\\
\end{Presented}
\vfill
\end{titlepage}

\section{Introduction}

Measurements of multiboson processes are a stringent test of the Standard Model (SM)
and have sensitivity to models of new physics.  
The LHC is a unique machine for studying multiboson processes because 
of its high center-of-mass energy and high luminosity.
The CMS experiment has measured a suite of multiboson
processes, which encompass signatures 
where two or more $W$ bosons, $Z$ bosons, or photons
are produced.
These processes have cross sections from a few hundred
picobarns to extremely rare processes of a few
femtobarns.  A summary of CMS measurements
of these processes as well as others can be found here\footnote{https://twiki.cern.ch/twiki/bin/view/CMSPublic/PhysicsResultsCombined}.


Many multiboson final states are produced through
triple or quartic gauge couplings.  These couplings
arise naturally from the non-Abelian nature of the
electroweak interaction.  The presence of new physics
may cause a modification of the these 
couplings which presents as an increase in the
measured cross section.  Such modifications
are generally enhanced in collisions having large
momentum transfer.  Therefore collisions at the LHC are ideal
for searching for new physics in the multiboson sector.  

\section{Parameterization of Anomalous Gauge Couplings}

Anomalous modifications of gauge couplings are 
parameterized using an Effective Field Theory (EFT)
which considers higher order modifications to the 
SM Lagrangian\cite{AGC},
\[
\mathcal{L}^{NP} = \mathcal{L}^{4 (SM)} + \frac{1}{\Lambda} \mathcal{L}^{5}+ \frac{1}{\Lambda^{2}} \mathcal{L}^{6}+ \frac{1}{\Lambda^{3}} \mathcal{L}^{7}+ \frac{1}{\Lambda^{4}} \mathcal{L}^{8}.
\]
The higher order terms are suppressed by 
a mass scale, $\Lambda$, for each 
additional power beyond the SM the term.
The terms of odd dimension are not considered as
they do not effect multiboson measurements or cause
significant anomalies such as lepton flavor violation
that are ruled out.
Therefore the dimension-6 and dimension-8 operators are considered.
Dimension-6 operators lead primarily to anomalous triple gauge
couplings (aTGC).  Anomalous quartic gauge couplings (aQGC)
arise from the dimension-8 operator.  Generally, if some
new physics exists, it is expected to be observed first as an aTGCs, 
but we also search for aQGC in the assumption that
no aTGCs exist.
In addition to the EFT parameterization, there are a number of 
other parameterizations that have been used in the past.
For anomalous neutral gauge couplings, a separate
parameterization is used as well.

\section{Experimental Measurements}

The CMS detector~\cite{CMSJINST} is designed to efficiently 
identify electrons, muons, photons, and hadronic jets giving a
high efficiency for selecting electroweak bosons.  The CMS 
detector is highly symmetric and covers pseudorapidity ranges
up to $|\eta| < 4.5$ so that escaping neutrinos can be identified 
from the missing transverse momentum ($E_{T}^{\text{miss}}$).
The results in this note are from the 2012 data taking period where approximately 20 fb$^{-1}$ of data
were recorded.  A summary of some results will be given
below.  More information can be found at the given references.

\subsection{ WW cross section measurements and limits on aTGC }

\begin{figure}
    \centering
    \subfloat[]{\includegraphics[width=0.45\textwidth]{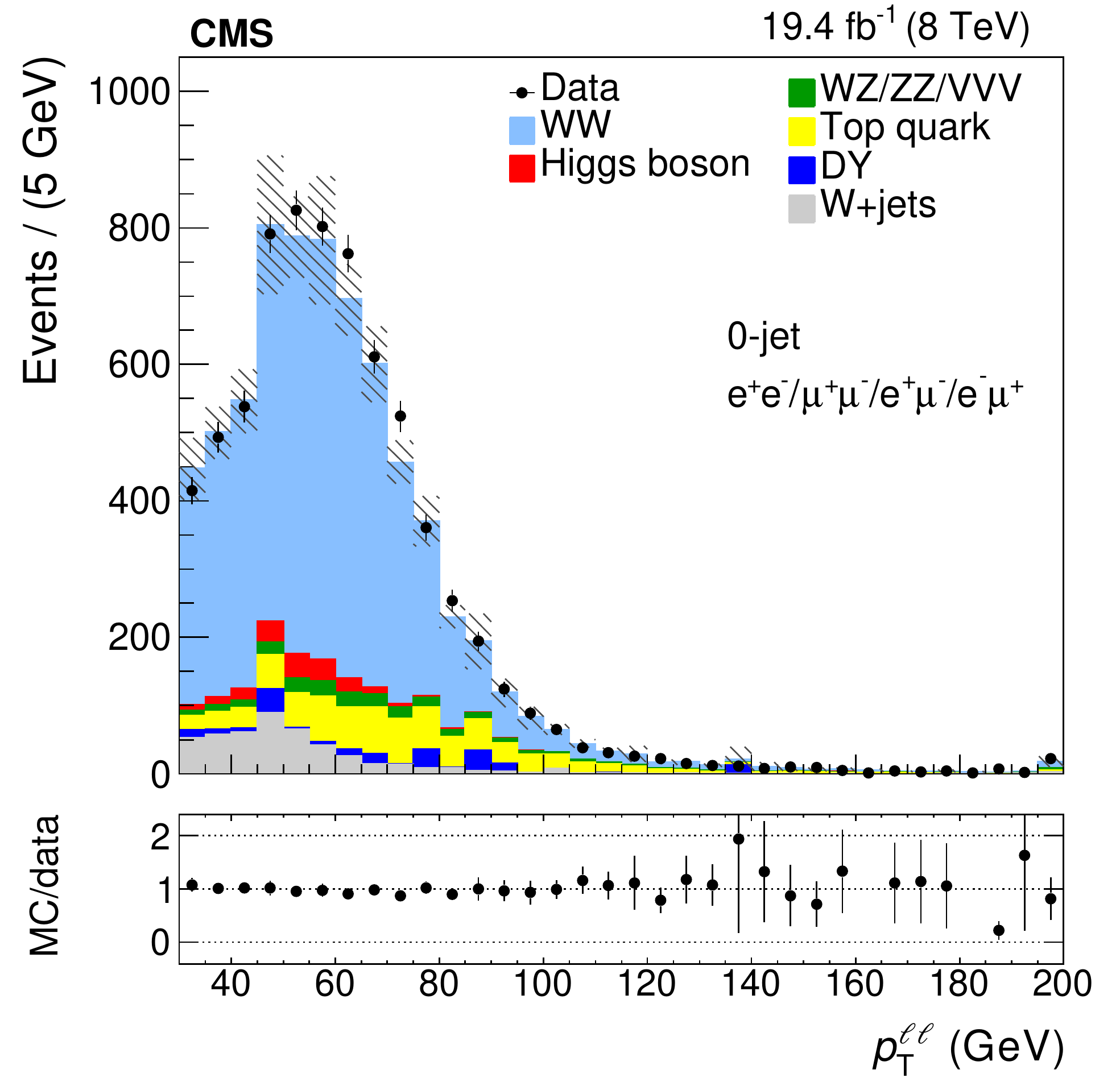}}
    \subfloat[]{\includegraphics[width=0.45\textwidth]{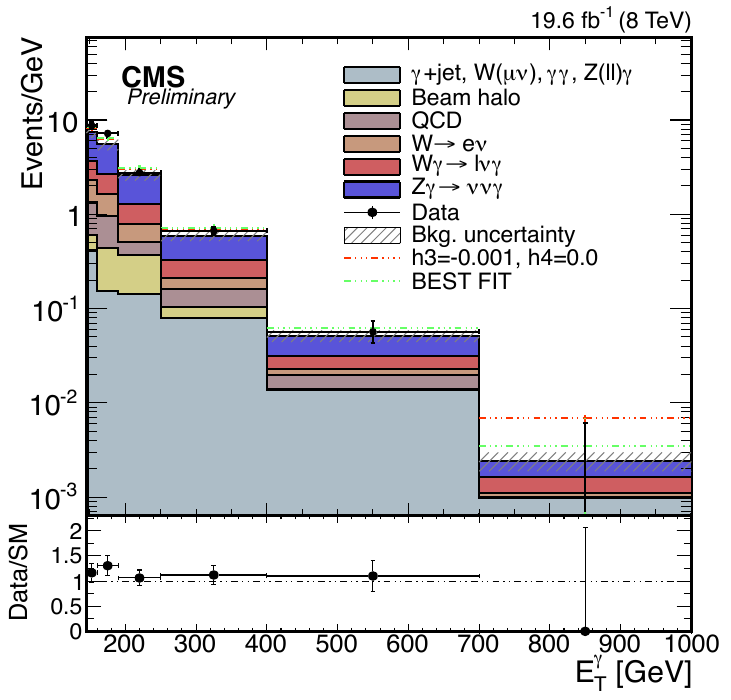}}

    \subfloat[]{\includegraphics[width=0.45\textwidth]{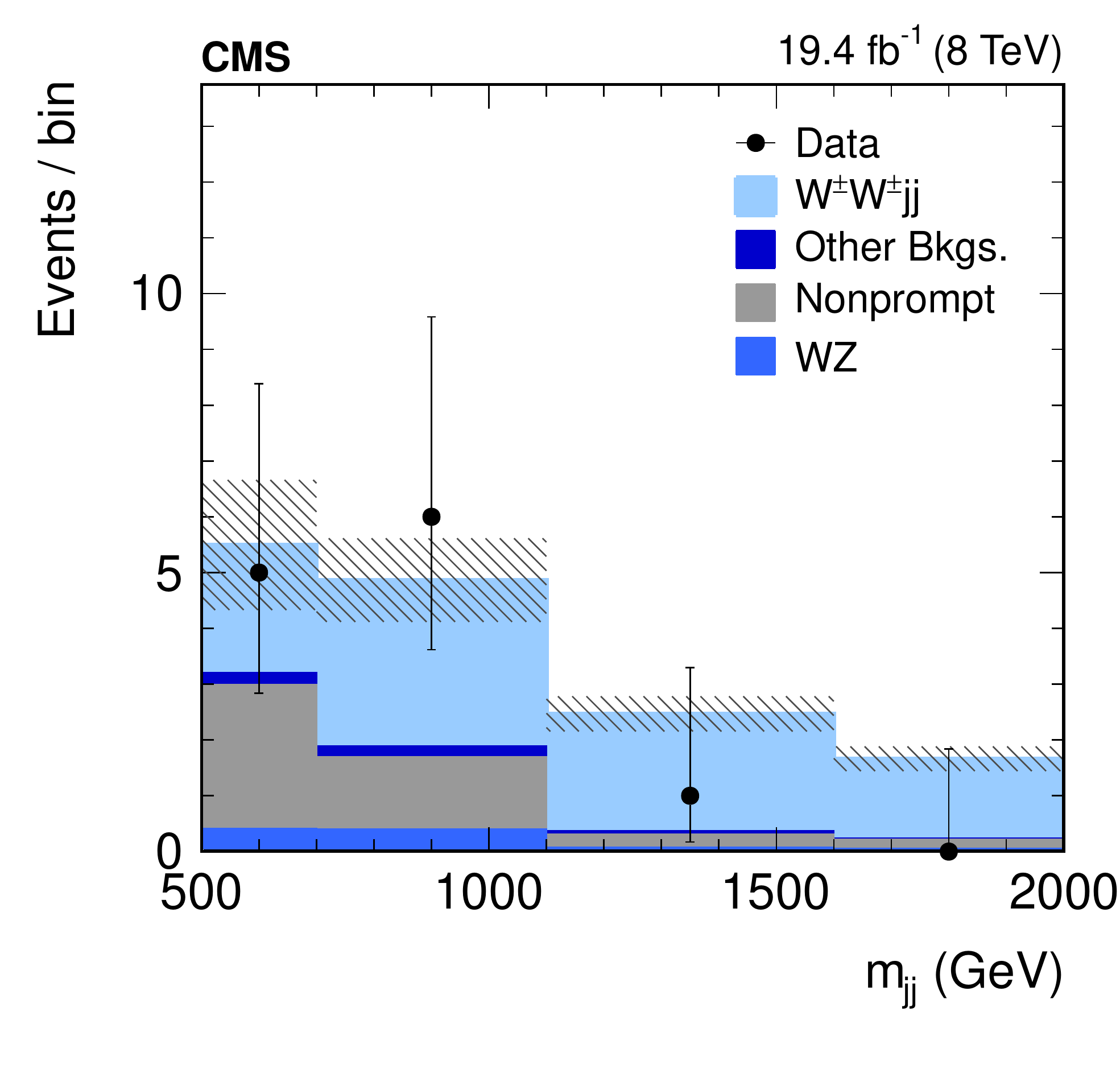}}
    \caption{\label{fig:all_results} 
        The dilepton $p_{T}$ spectrum for signal $WW$ events having no reconstructed
        jets in the event (a).  Photon $E_{T}$ spectrum
        for the $Z\gamma$ analysis (b).  The predictions
        show the expectation from some models
        of anomalous gauge couplings.
        Dilepton mass spectrum for events that pass the 
        exclusive vertex selection and have
        one electron and one muon (c).   
    Dijet mass spectrum for events that have
        two same-sign leptons and two jets with
        large pseudorapidity separation (d).
        }
\end{figure}

\begin{figure}
\centering
    \subfloat[]{\includegraphics[width=0.45\textwidth]{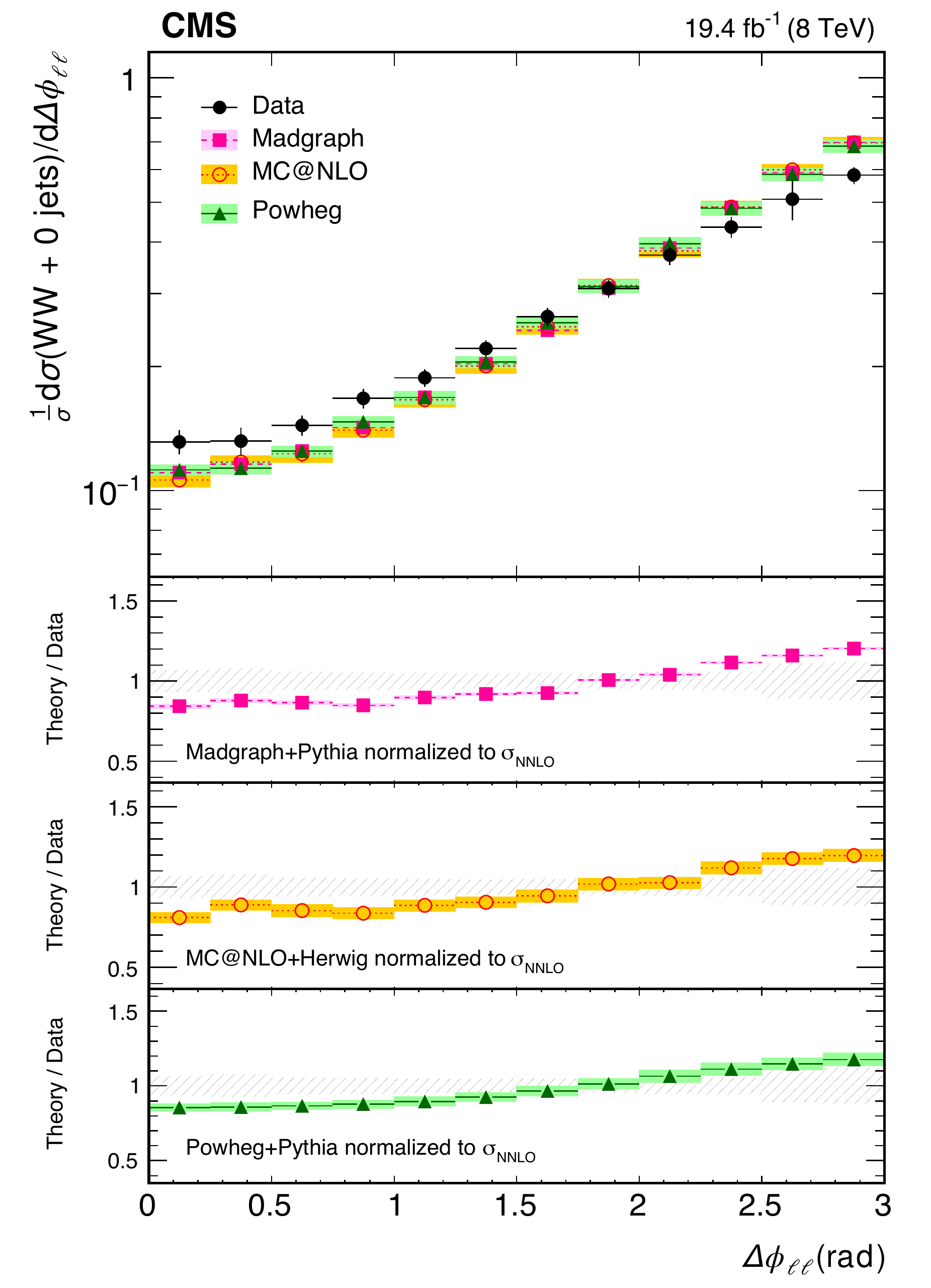}}
    \subfloat[]{\includegraphics[width=0.45\textwidth]{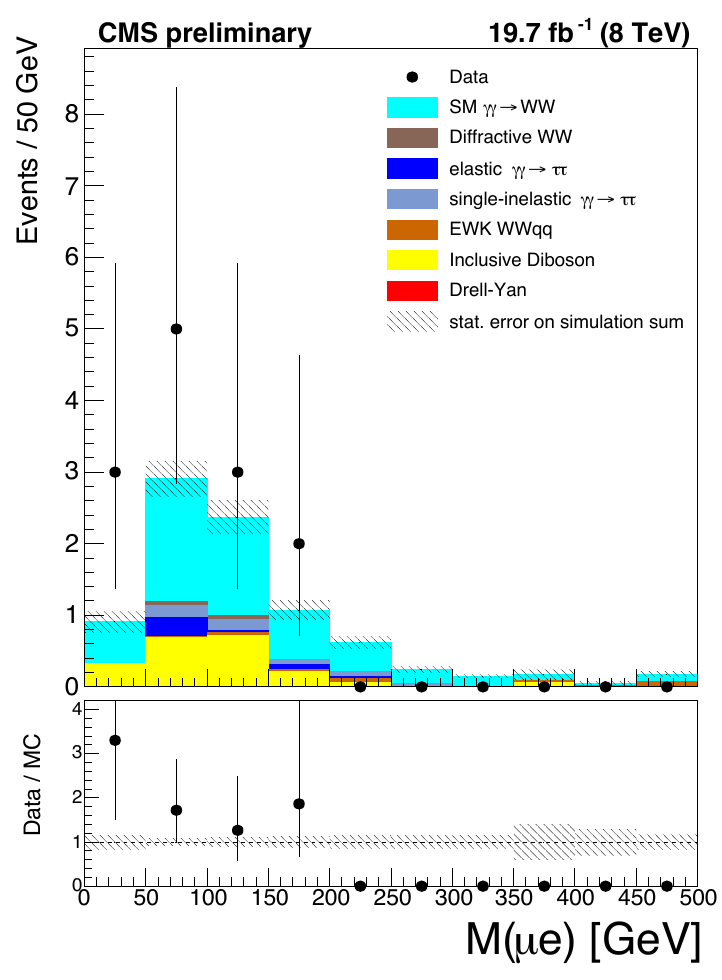}}
\caption{\label{fig:ww_results} Differential cross section as a function of the
            azimuthal separation of the leptons.  The observed events are unfolded to the
            parton level and compared to a number of NLO theoretical predictions.
}
\end{figure}

The production of opposite-sign $W$ boson pairs is a high
rate process at the LHC having a theoretical cross section of
$59.8 \pm 2.2$ pb calculated at NNLO~\cite{WWNNLO}.
The $WW$ process is identified from fully leptonic decays
in the di-electron, di-muon, or electron-muon final state~\cite{WW8TEV}.
Large $E_{T}^{\text{miss}}$ is required to reduce backgrounds
by identifying the escaping neutrinos. 
To further reduce backgrounds from Drell-Yan in the same-flavor channels 
an MVA-based discriminator is used, and to reduce backgrounds
from $t\bar{t}$ and single top, events are rejected if a jet is tagged as
originating from a b-hadron.  
Data-driven methods are used to estimate the $W$, Drell-Yan, and top quark backgrounds
and other, smaller backgrounds are estimated using Monte Carlo simulation.
Figure~\ref{fig:all_results} (a) shows the dilepton $p_{T}$ spectrum for selected
events having no reconstructed jets.
The measured cross section is 60.1 $\pm$ 0.9 (stat) $\pm$ 3.2 (exp) $\pm$ 3.2 (theory) $\pm$ 1.6 (lumi) pb,
consistent with the NNLO theoretical prediction.  
Differential cross sections of a number of quantities are measured.
The differential cross sections are unfolded to the parton level
and are compared to a number of theoretical predictions as shown
in Figure~\ref{fig:ww_results} (a).
In addition, improved limits are placed on aTGC couplings.

\subsection{ Measurement of $Z\gamma$ in the $Z\rightarrow \nu\nu$ final state}

The production of $Z\gamma$ is another high-rate process at the LHC.  
This measurement takes advantage of the relatively higher 
branching fraction of the $Z$ boson decay to neutrinos 
by selecting one photon with large transverse momentum
and large $E_{T}^{\text{miss}}$ directed oppositely to the photon~\cite{Zgamma}.
Backgrounds from, $W$ events where the decay electron
mimics a photon, $\gamma$ + jet, and instrumental sources
are estimated using data-driven methods.  The background 
from $W\gamma$ events where the decay lepton is lost
is estimated using simulation.
The photon $E_{T}$ spectrum with the signal and
background predictions is shown in Figure~\ref{fig:all_results} (b).
The measured cross section in the fiducial region
where the photon has transverse momentum greater
than 145 GeV is 52.7 $\pm$ 2.1 (stat) $\pm$ 6.4 (syst) $\pm$ 1.4 (lumi) fb,
consistent with the theoretical prediction of 50.0 $\pm$ 2.4 fb.
In addition stringent limits are set on anomalous $ZZ\gamma$ and
$Z\gamma\gamma$ neutral gauge couplings.

\subsection{ Measurement of $WW$ photoproduction }

This measurement identifies events where
the incoming protons radiate photons, producing
two opposite-sign $W$ bosons.  The interaction can occur through
triple or quartic gauge couplings.  In these events
the protons receive only a small transverse momentum
and are lost out of the acceptance of the detector.
Such events are identified from the leptonic decays of the
$W$ bosons with no additional activity from the interaction 
vertex~\cite{ggWW}. 
Backgrounds in the selection are estimated with Monte Carlo simulations
with systematic uncertainties determined from dedicated control regions.
Figure~\ref{fig:ww_results} (b) shows the dilepton mass spectrum for
events passing the event selection and having one electron and one muon.
The measured cross section is 12.3$^{+5.5}_{-4.4}$ fb
with a theoretical cross section of 6.9 $\pm$ 0.6 fb.  The 
significance of the observed events over background is 3.6$\sigma$.
In addition, limits are placed on aTGCs and aQGCs.

\subsection{ Measurement of same-sign $WW$ production with forward jets}

This measurement searches for the unique signature of tow
same-sign $W$ bosons in vector boson scattering (VBS) events. 
In these events both incoming protons radiate $W$ bosons which
scatter in a quartic gauge vertex.  The event signature is 
two leptons having the same charge, and two forward jets~\cite{WWSS}.
The signal is enhanced by requiring the jets to
have large mass and large pseudorapidity separation.
Figure~\ref{fig:all_results} (c) shows the dijet mass distribution
fro signal events.
Backgrounds from non-prompt leptons are estimated using 
a data-driven method and the remaining backgrounds are
estimated using simulation.  The signal is observed
with a significance of $2.0\sigma$ and stringent
limits are placed on aQGC.

\subsection{Conclusions}

The large integrated luminosity delivered by the LHC
in the 2012 data set allows for the measurement 
of many multiboson processes.  Precision measurements
of multiboson processes test the electroweak gauge sector.
With NNLO theoretical predictions available for many
multiboson processes precision measurements become more 
powerful.  An Effective Field Theory is used to
test for the presence of possible new
physics and limits are set on the coupling parameters.
Many multiboson measurements have been made
with CMS; a selection of measurements that are representative
of the SM physics program was given. 

\bibliography{proceedings}{}
\bibliographystyle{unsrt}

\end{document}